\documentclass[showpacs,aps,graphicx,twocolumn]{revtex4}
\usepackage{amsmath}
\usepackage{graphicx}

\begin{document}


\title{Heralded high-efficiency quantum repeater with atomic ensembles assisted by faithful single-photon transmission\footnote{Published in
Sci. Rep. \textbf{5}, 15610
(2015)}}

\author{Tao Li$^1$  and  Fu-Guo Deng$^{1,2}$\footnote{Corresponding author:
fgdeng@bnu.edu.cn} }
\address{$^1$ Department of Physics, Applied Optics Beijing Area Major Laboratory, Beijing normal University, Beijing 100875, China\\
$^2$ State Key Laboratory of Networking and Switching Technology,
Beijing University of Posts and Telecommunications, Beijing 100876,
China}
\date{\today }

\begin{abstract}
Quantum repeater is one of the important building blocks for long
distance quantum communication network. The previous quantum
repeaters based on atomic ensembles and linear optical elements can
only be performed with a maximal success probability of 1/2 during
the entanglement creation and entanglement swapping procedures.
Meanwhile, the polarization noise during the entanglement
distribution process is harmful to the entangled channel created.
Here we introduce a general interface between a polarized photon and
an atomic ensemble trapped in a single-sided optical cavity, and
with which we propose a high-efficiency quantum repeater protocol in
which the robust entanglement distribution is accomplished by the
stable spatial-temporal entanglement and it can in principle create
the deterministic entanglement between neighboring atomic ensembles
in a heralded way as a result of cavity quantum electrodynamics.
Meanwhile, the simplified parity check gate makes the entanglement
swapping be completed with unity efficiency, other than 1/2 with
linear optics. We detail the performance of our protocol with
current experimental parameters and show its robustness to the
imperfections, i.e., detuning and coupling variation, involved in
the reflection process. These good features make it a useful
building block in long distance quantum communication.
\end{abstract}


\maketitle

\section{Introduction}

Quantum mechanics provides some interesting ways for communicating
information  securely between remote parties
\cite{teleportation,QKD,QKD2,QSDC1,QSDC2}. However, in practice the
quantum channels such as optical fibers are noisy and lossy
\cite{Qcryptography}. The transmission loss and the decoherence of
photon systems  increase exponentially with the distance, which
makes it extremely hard to perform a long-distance quantum
communication directly. To overcome this limitation, Briegel
\emph{et al}. \cite{Qrepeater} proposed a noise-tolerant quantum
repeater protocol in 1998. The channel between the two remote
parties \emph{A} and \emph{B} is divided into smaller segments by
several nodes, the neighboring nodes can be entangled efficiently by
the indirect interaction through flying qubits, and the entanglement
between non-neighboring nodes is implemented by quantum entanglement
swapping, which can be cascaded to create the entanglement between
the terminate nodes \emph{A} and \emph{B}.

The implementation of quantum repeaters is compatible with different
physical setups assisted by cavity quantum electrodynamics, such as
nitrogen vacancy  centers in diamonds \cite{high-frep1}, spins in
quantum dots
\cite{qdrepeater1,qdrepeater12,qdrepeater2,qdrepeater3}, single
trapped ions or atoms \cite{qdions,atomrepeater2}.  However, the
most  widely  known approach for quantum repeaters is based on
atomic ensembles \cite{reviewqr} due to the collective enhancement
effect \cite{colle1}. In a seminal paper by Duan \emph{et al.}
\cite{DLCZ}, the atomic ensemble is utilized to act as a local
memory node. The heralded collective spin-wave entanglement between
the neighboring nodes is established by the detection of a single
Stokes photon, emitted indistinguishably from either of the two
memory nodes via a Raman scattering process. However, due to the low
probability of Stokes photon emission required in the
Duan-Lukin-Cirac-Zoller (DLCZ) proposal \cite{DLCZ}, the parties can
hardly establish the entanglement efficiently for quantum
entanglement swapping.  In order  to  improve  the  success
probability, photon-pair sources and multimode memories are used to
construct a temporal multi-mode modification \cite{QRsimontimemul},
and then the schemes based on the single-photon sources
\cite{QRsingleSangouard} and spatial multiple modes
\cite{QRsimonspatialmul} are developed. Besides these protocols
based on Mach-Zehnder-type interference, Zhao \emph{et al.}
\cite{QRzhao,QRchen} proposed a robust quantum repeater protocol
based on two-photon Hong-Ou-Mandel-type  interference, which relaxes
the long-distance  stability requirements and suppresses the vacuum
component to  a constant item. Subsequently, the single-photon
sources are embedded to improve the performance of robust quantum
repeaters \cite{QRdoubleSangouard,QRlingw,QRxiongsj}. In addition,
Rydberg blockade effect \cite{Rydbergorig} is used to perform
controlled-NOT  gate between the two atomic ensembles in the middle
node \cite{QRhanzhao,QRzhaohan}, which makes the quantum
entanglement swapping operation be performed deterministically.

Since the two-photon interference is performed with the polarization
degree of freedom (DOF) of the photons \cite{QRzhao,QRchen}, which
is incident to be influenced by the thermal fluctuation, vibration,
and the imperfection of the fiber \cite{Faithful}, the fidelity of
the entanglement created between the neighboring nodes will be
decreased when the photons are transmitted directly
\cite{Qcryptography,Qrepeater}. In other words, the more the overlap
of the initial photon state used in the two-photon interference is,
the higher the fidelity of the entanglement created is. Following
the idea of Zhao's protocol \cite{QRzhao}, quantum repeaters immune
to the rotational polarization noise are proposed with the time-bin
photonic state \cite{QRgaom} and the antisymmetric Bell state
\cite{QRdfs} $|\Psi^-\rangle=(|HV\rangle-|VH\rangle)/\sqrt{2}$,
respectively. When the noise on the two orthogonal polarized photon
states is independent, Zhang \emph{et al.} \cite{QRzhangbb} utilized
the faithful transmission of polarization photons \cite{Faithful} to
surmount the collective noise. In the ideal case, the two-fold
coincidence detection in the central node can successfully get the
stationary qubits entangled  maximally in a heralded way. Apart from
this type of entanglement distribution, Kalamidas \cite{Faithfulerr}
proposed an error-free entanglement distribution protocol in the
linear optical repeater. An entangled photon source is placed at the
center node, and the entangled photons transmitted to neighboring
nodes are encoded with their time-bin DOF. With two fast Pockels
cells (PCs), the entanglement distribution can be performed with a
high efficiency when the polarization-flip-error noise is relatively
small.

In a recent work, Mei \emph{et al.} \cite{Meirepeater} built a
controlled-phase-flip (CPF) gate between a flying photon  and an
atomic ensemble embedded in an optical cavity, and constructed a
quantum repeater protocol, following some ideas in the original DLCZ
scheme \cite{DLCZ}. In 2012, Brion \emph{et al.} \cite{QRBrion}
constituted a quantum repeater protocol with Rydberg blocked atomic
ensembles in fiber-coupled cavities via collective laser
manipulations of the ensembles and photon transmission. Besides,
Wang \emph{et al.} \cite{WangtjOEA} proposed a one-step
hyperentanglement distillation and amplification proposal, and Zhou
and Sheng \cite{ShengLPLA} designed a recyclable protocol for the
single-photon entanglement amplification, which are quite useful to
the high dimensional or multiple DOFs optical quantum repeater.

In this paper, we give a general interface between a polarized
photon and an atomic ensemble trapped in a single-sided optical
cavity. Besides, we show that a deterministic faithful entanglement
distribution in a quantum repeater can be implemented with the
time-bin photonic state when two identical fibers act as the
channels of different spatial DOFs of the photons. Interestingly, it
does not require fast PCs and the time-slot discriminator
\cite{Faithful,QRgaom,QRdfs,QRzhangbb,Faithfulerr} is not needed
anymore. By using the input-output process of a single photon based
on our  general interface, the entanglement between the neighboring
atom ensembles can be created in a heralded way, without any
classical communication after the clicks of the photon detectors,
and the quantum swapping can be implemented with almost unitary
success probability by a simplified parity-check gate (PCG) between
two ensembles, other than $1/2$ with linear optics. We analyze  the
performance of our high-efficiency quantum repeater protocol with
current experimental parameters and show its robustness to the
imperfections involved in the reflection process. These good
features  will make it a useful building block in long-distance
quantum communication in future.

\section{Results}

\subsection{A general interface between a polarized photon and an atomic ensemble.}

The elementary node in our quantum repeater protocol includes an
ensemble with $N$ cold atoms trapped in a single-sided optical
cavity \cite{Meirepeater,QRBrion}. The atom has a four-level
internal structure and its relevant levels are shown in Fig. 1. The
two hyperfine ground states are denoted as $|g\rangle$ and
$|g_h\rangle$. The excited state $|e\rangle$ and the Rydberg state
$|r\rangle$ are two auxiliary states. The $|h\rangle$ polarized
cavity mode $a_h$ couples to the transition between $|g_h\rangle$
and $|e\rangle$. Initially, all of the atoms are pumped to the state
$|g\rangle$. With the help of the Rydberg state $|r\rangle$, one can
efficiently perform an arbitrary operation between the ground state
$|G\rangle=|g_1,\dots,g_j,\dots,g_{_N}\rangle$ and the single
collective spin-wave excitation state \cite{DLCZ}
$|S\rangle=\frac{1}{\sqrt{N}}\sum_{j=1}^{_N}|g_1,\dots,g_{h_j},\dots,g_{_N}\rangle$
via collective laser manipulations of the ensembles
\cite{Meirepeater,QRBrion,Rydbergcollectiveencodeing}. The single
collective  excited state
$|E\rangle=\frac{1}{\sqrt{N}}\sum_{j=1}^{_N}|g_1,\dots,e_{_j},\dots,g_{_N}\rangle$.
When the  Rydberg blockade shift is of the scale $ 2\pi\times
100MHz$, the transition  between $|G\rangle$  and $|S\rangle$ can be
completed with an effective coupling strength $2\pi \times 1MHz$ and
the probability of nonexcited and doubly excited errors
\cite{Spinerror1} is about $10^{-3}$-$10^{-4}$. Recently, rotations
along axes  $R_x$, $R_y$, and $R_z$ of a spin-wave excitation with
an average fidelity of $99\%$ are achieved in ${}^{87}Rb$ atomic
ensembles and they are implemented by making use of stimulated Raman
transition and controlled Larmor procession \cite{Spinerror2}. In
other words,  the high-efficiency single qubit rotations of the
atomic ensemble can be implemented faithfully.

\begin{figure}[htbp]             
\centering\includegraphics[width=8 cm]{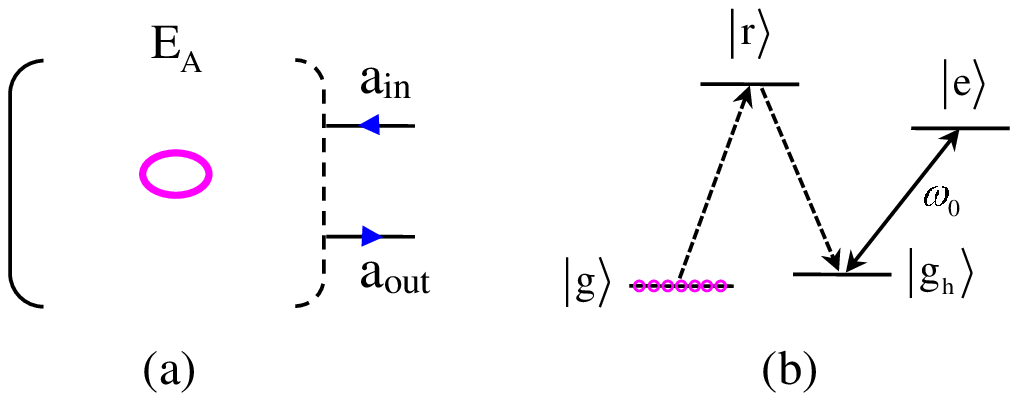} \caption{ (Color
online)  (a) Schematic diagram for a  single-side  cavity coupled to
an atomic ensemble system. (b) Atomic level
structure.\label{figure1}}
\end{figure}

Let us consider an $|h\rangle$ polarized input photon with the
frequency $\omega$, which is nearly resonant to the cavity mode
$a_{_h}$ with the frequency $\omega_c$. The coupling rate between
the cavity and the input photon can be taken to be a real constant
$\sqrt{\frac{\kappa}{2\pi}}$ when the detuning $|\delta'|= |\omega
-\omega_c|$ is far less than the cavity decay rate $\kappa$
($|\delta'|\ll\kappa$)
\cite{quantumoptics,neutralcomp1,neutralcomp2}. The Hamiltonian of
the whole system, in the frame rotating with respect to the cavity
frequency $\omega_c$, is ($\hbar=1$) \cite{quantumoptics}
\begin{eqnarray}                  
\hat{H}_s
\!\!&=&\!\!\sum_{j=1}^{N}\left[\left(\Delta-i\frac{\gamma_{e_{j}}}{2}\right)\!\hat{\sigma}_{e_{j}{e_{j}}}
+ig_j\!\left(\hat{a}_{_h}\hat{\sigma}_{e_{j}{s_{j}}}-\hat{a}_{_h}^{\dagger}\hat{\sigma}_{s_{j}{e_{j}}}\right)\right] \nonumber\\
&&+i\sqrt{\frac{\kappa}{2\pi}}\int{}\!d\delta'\!\left[\hat{b}^{\dagger}(\delta')\hat{a}_{_h}\!-\!\hat{b}(\delta')\hat{a}_{_h}^{\dagger}\right]\!+\!\!
\int\!
d\delta'\hat{b}^{\dagger}(\delta')\hat{b}(\delta'),\nonumber\\
 \label{hami}
\end{eqnarray}
where $\hat{a}$ and  $\hat{b}$ are the operators of the cavity mode
and the input photon with the properties
$[\hat{a},\hat{a}^{\dagger}]=1$ and
$[\hat{b}(\delta'),\hat{b}^{\dagger}(\delta'')]=\delta(\delta'-\delta'')$,
respectively. $\Delta=\omega_0-\omega_c$ is the detuning between the
cavity mode frequency $\omega_c$  and the dipole transition
frequency $\omega_0$,
$\hat{\sigma}_{e_{j}{e_{j}}}=|e_{j}\rangle\langle e_{j}|$, and
$\hat{\sigma}_{e_{j}{s_{j}}}=|e_{j}\rangle\langle s_{j}|$.
$\gamma_{e_{j}}$ represents the spontaneous emission rate of the
excited state $|e_{j}\rangle$, while $g_{j}$ denotes  the coupling
strength  between  the \emph{j-th} atom transition and the cavity
mode $\hat{a}_{_h}$. Here and after,  we  assume $g_j = g$ and
$\gamma_{e_{j}} = \gamma$ for simplicity .

With the Hamiltonian $\hat{H}_s$ shown in Eq.(\ref{hami}), the
Heisenberg-Langevin equations of motion for cavity $\hat{a}_h$ and
the atomic operator
 $\hat{ \sigma}_{-}=|S\rangle{}\langle{}E|$ taking into account the atomic excited state decay $\gamma$ can be detailed as \cite{quantumoptics}
\begin{eqnarray}   
\begin{split}
\frac{d \hat{a}_h}{d t}\;=\;&-\left(
i\omega_c+\frac{\kappa}{2}\right)\hat{a}_h-ig\hat{\sigma}_{-}
-\sqrt{\kappa}\,\hat{a}_{in},\\
\frac{d \hat{\sigma}_{-}}{d t}\;=\;&-\left(
i\omega_{0}+\frac{\gamma}{2}
\right)\hat{\sigma}_{-}+ig\hat{\sigma}_{z}\hat{a}_h+\sqrt{\gamma}\,\hat{\sigma}_{z}\hat{N}.
\label{Heisenberg}
\end{split}
\end{eqnarray}
Here the Pauli operator
$\hat{\sigma}_{z}=|E\rangle{}\langle{}E|-|S\rangle{}\langle{}S|$,
while $\hat{N}$ is corresponding to the vacuum noise  field that
helps to preserve the desired commutation relations for the atomic
operator. Along with the standard cavity input-output relation
$\hat{a}_{out}=\hat{a}_{in}+\sqrt{\kappa}\,\hat{a}_h$, one can
obtain the reflection and noise coefficients $r(\delta')$ and
$n(\delta')$ in the weak excitation approximation where  the
ensemble is hardly in the state $|E\rangle$ but predominantly in
$|S\rangle$, that is,
\begin{eqnarray}                  
\label{r}
\begin{split}
r{(\delta')}\;=\;&\frac{(\delta'-i\kappa/2)(\Delta'+i\gamma/2)-g^2}{(\delta'+i\kappa/2)(\Delta'+i\gamma/2)-g^2},\\
n{(\delta')}\;=\;&\frac{ig\sqrt{\kappa\gamma}}{(\delta'+i\kappa/2)(\Delta'+i\gamma/2)-g^2},
\end{split}
\end{eqnarray}
where $\Delta'=\omega-\omega_0$ represents the frequency detuning
between the input photon and the dipole transition.
$|r{(\delta')}|^2+|n{(\delta')}|^2=1$  means that when the noise
field is considered, the energy is conserved during the input-output
process of  the single-sided cavity.

\begin{figure}[!h]
\begin{center}
\includegraphics[width=8.0 cm]{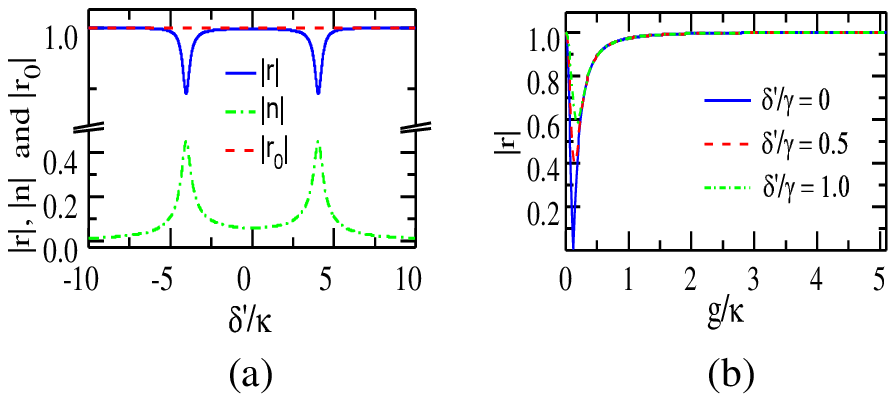}
\caption{(Color online)  (a)  $|r|$, $|n|$ and $|r_0|$ vs the scaled
detuning $\delta'/\kappa$, with the scaled coupling rate
$g/\kappa=4.0566$ and $\gamma/\kappa=0.0566$ \cite{gvarious}.  (b)
$|r|$ vs the scaled coupling rate $g/\kappa$ with detuning
$\delta'/\gamma=0,0.5,$ and $1$.}\label{Fig2}
\end{center}
\end{figure}

If the  atomic  ensemble in the cavity is initialized to be the
state $|G\rangle$, it does not interact with the cavity mode (i.e.,
$g=0$). The input $|h\rangle$ polarized probe photon feels an empty
cavity and will be reflected by the cavity directly. Now, the
reflection coefficient can be simplified to be \cite{quantumoptics}
\begin{eqnarray}              
\label{r0}
r_{_0}{(\delta')}=\frac{\delta'-i\kappa/2}{\delta'+i\kappa/2}.
\end{eqnarray}
Note that the detuning is small $|\delta'|\ll \kappa$, the pulse
bandwidth  is much less than the cavity decay rate $\kappa$. If the
strong coupling condition $\gamma\kappa/4\ll{}g^2$ is achieved,  one
can get the input probe photon totally reflected with
$r_{_0}{(\delta')}\simeq-1$ or $r{(\delta')}\simeq1$, shown in Fig.
2. The absolute phase shifts versus the scaled detuning are shown in
Fig. 3.

\begin{figure}[!h]
\begin{center}
\includegraphics[width=8.0 cm]{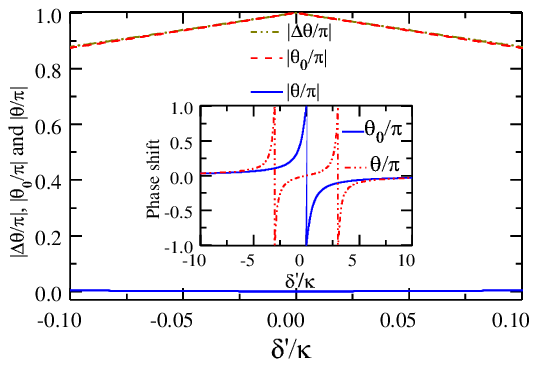}
\caption{(Color online) The absolute phase shifts vs the scaled
detuning. The dashed and dashed-dot lines show the absolute phase
shifts $|\theta_0/\pi|$ and $|\theta/\pi|$ that the reflected photon
gets, with the ensemble in $|G\rangle$ and $|S\rangle$,
respectively. The solid line represents the absolute value of the
phase shifts difference $|\Delta\theta/\pi|$ =
$|\theta_0/\pi-\theta/\pi|$. The inset shows the phase shifts vs the
scaled detuning $\theta_0/\pi$ and $\theta/\pi$ that the reflected
photon gets,  with the ensemble in $|G\rangle$ and $|S\rangle$,
respectively.}\label{Fig3}
\end{center}
\end{figure}

\subsection{Hybrid CPF gate on a photon-atomic-ensemble  system  and PCG on a two-atomic-ensemble system.}

The principle of our CPF gate on a hybrid quantum system composed of
a photon $p$ and an atomic ensemble  $E_A$ is shown in Fig. 4,
following some ideas in previous works
\cite{Meirepeater,neutralcomp1,neutralcomp2}. Suppose that the
photon $p$ is in the state
$|\varphi_p\rangle=\mu|h\rangle+\nu|v\rangle$ ($|\mu|^2+|\nu|^2=1$)
and the ensemble $E_{_A}$ is in the state
$|\phi_{_A}\rangle=\mu'|G\rangle+\nu'|S\rangle$ ($
|\mu'|^2+|\nu'|^2=1$).  The $|h\rangle$ polarized component of the
photon $p$ transmits the polarization beam splitter (PBS) and  then
be reflected by the cavity, while the $|v\rangle$ polarized
component is reflected by the mirror $M$.  The optical pathes of the
$|h\rangle$ and $|v\rangle$ components are adjusted to be equal and
they will be combined again at the PBS with an extra $\pi$ phase
shift on the $|h\rangle$ component if the ensemble is in the state
$|G\rangle$. This process can be described as
\begin{eqnarray}     
\begin{split}
|\phi_{_A}\rangle \otimes |\varphi_p\rangle\rightarrow &\mu'|G\rangle\otimes(-\mu|h\rangle+\nu|v\rangle)\\
&+\nu'|S\rangle\otimes(\mu|h\rangle+\nu|v\rangle).
\end{split}
\end{eqnarray}
That is to say, the setup in Fig. 4(a) can be used to accomplish a
CPF gate  on the atomic ensemble $E_{_A}$ and the photon $p$.

\begin{figure}[!h]
\begin{center}
\includegraphics[width=8.0 cm,angle=0]{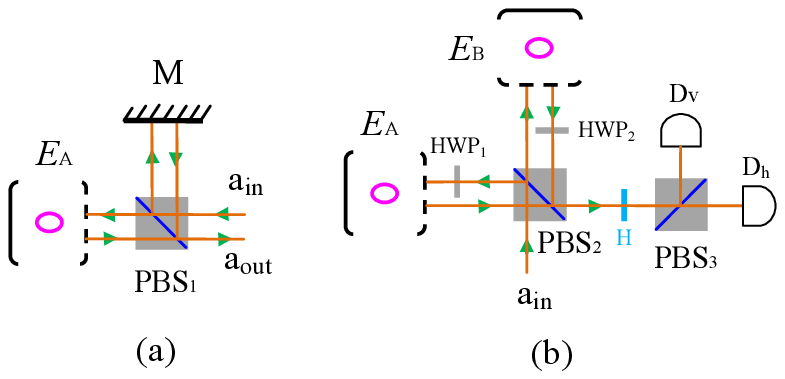}
\caption{(Color online) Schematic setup for implementing a CPF gate
and a parity-check gate (PCG). $M$ stands for a mirror and the PBS
transmits the $|h\rangle$ polarized photon and reflects the
$|v\rangle$ component. HWP$_1$ and HWP$_2$ are half wave plates
performing the bit-flip operation while H represent a Hadamard
rotation.} \label{Fig4}
\end{center}
\end{figure}

The schematic  diagram of our PCG on two atomic ensembles $E_{_A}$
and $E_{_B}$ is shown in Fig. 4(b). Let us assume that  $E_A$ and
$E_B$ are initially in the states
$|\phi_{i}\rangle=\mu_{i}|G\rangle_{i}+\nu_{i}|S\rangle_{i}$ (
$|\mu_{i}|^2+ |\nu_{i}|^2=1$ and  $i=A$, $B$).
 One can input a polarized photon \emph{p} in the state
$|\varphi_p\rangle=\frac{1}{\sqrt{2}} (|h\rangle+|v\rangle)$ into
the import of the setup.  HWP$_1$ (HWP$_2$) is used to perform the
bit-flip operation $|h\rangle \leftrightarrow|v\rangle$ on the
photon $p$ by using a half-wave plate (HWP) with its axis at $\pi/4$
with respect to the horizontal direction.  After the two components
of $p$ are reflected by the two cavities, they combine with each
other at PBS$_2$.  The state of the system composed of the two atom
ensembles and the photon evolves to be
\begin{eqnarray}         
|\Phi\rangle_{p_{_{AB}}}\!\!\!&=&\!\!\!\frac{1}{\sqrt{2}}[
|h\rangle\!\otimes\!(-\mu_{_{A}}|G\rangle_{_{A}}\!+\!\nu_{_{A}}|S\rangle_{_{A}})(\mu_{_{B}}|G\rangle_{_{B}}\!+\!\nu_{_{B}}|S\rangle_{_{B}}) \nonumber\\
&&+
|v\rangle\!\otimes\!(\mu_{_{A}}|G\rangle_{_{A}}\!+\!\nu_{_{A}}|S\rangle_{_{A}})(-\mu_{_{B}}|G\rangle_{_{B}}\!+\!\nu_{_{B}}|S\rangle_{_{B}})].\nonumber\\
\end{eqnarray}
And then,   another HWP names $H$ whose axis is placed at $\pi/8$
is used to perform a Hadamard rotations
$|h\rangle\leftrightarrow1/\sqrt{2}(|h\rangle+|v\rangle)$ and
$|v\rangle\leftrightarrow1/\sqrt{2}(|h\rangle-|v\rangle)$ on the
photon. The state of the system becomes
\begin{eqnarray}     
\label{phipab}
|\Phi\rangle'_{p_{AB}}\!\!&=&\!\!
|h\rangle\otimes(\nu_{_{A}}\nu_{_{B}}|S\rangle_{_{A}}|S\rangle_{_{B}}
- \mu_{_{A}}\mu_{_{B}}|G\rangle_{_{A}}|G\rangle_{_{B}}) \nonumber\\
&&\!+
|v\rangle\otimes(\nu_{_{A}}\mu_{_{B}}|S\rangle_{_{A}}|G\rangle_{_{B}} - \mu_{_{A}}\nu_{_{B}}|G\rangle_{_{A}}|S\rangle_{_{B}}).\nonumber\\
\end{eqnarray}
After the photon is measured with PBS$_3$ and two single-photon
detectors, the parity of $E_A$  and $E_B$ can be determined. In
detail, if the  photon is  in the state $|h\rangle$, the two
ensembles $E_A$ and $E_B$ have an even parity. If the  photon is in
$|v\rangle$,  $E_A$ and $E_B$ have an odd parity. With an effective
input-output process of a single photon, one can efficiently
complete the  PCG on two atomic ensembles.

\begin{figure}[!h]
\begin{center}
\includegraphics[width=8 cm]{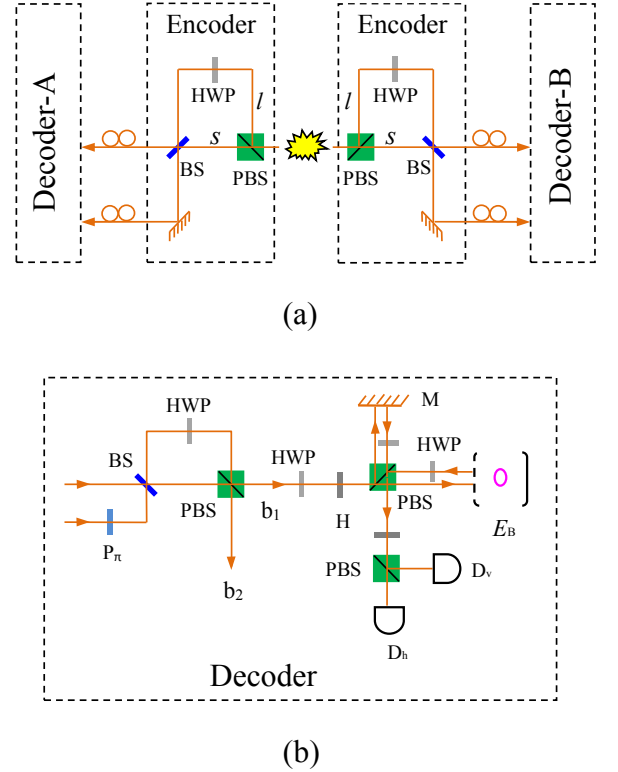}
\caption{(Color online) Schematic setup for entanglement
distribution. p$_\pi$ is a $\pi$ phase shifter.} \label{Fig5}
\end{center}
\end{figure}

\subsection{Entanglement distribution with faithful single-photon transmission.}

Suppose that  there is an entanglement source which is placed at a
central station between two neighboring nodes, say Alice and Bob.
The source produces a two-photon polarization-entangled Bell state
$|\Psi^+\rangle_{ab}=1/\sqrt{2}(|h\rangle_a|v\rangle_b+|v\rangle_a|h\rangle_b)$.
Here  the subscripts  $a$   and $b$  denote  the  photons  sent  to
Alice and Bob, respectively. As shown in Fig. 5 (a), the photons
\emph{a} and \emph{b} will pass through an encoder in each side
before they enter the noisy channels. The encoder is made up of a
PBS, an  HWP,  and a beam splitter (BS).  Here  BS is used for a
Hadamard rotation on the spatial DOF of the photon, i.e.,
$|u\rangle\leftrightarrow \frac{1}{\sqrt{2}} (|u\rangle+|d\rangle)$
and $|d\rangle\leftrightarrow \frac{1}{\sqrt{2}}
 (|u\rangle-|d\rangle)$, where $|u\rangle$ and $|d\rangle$ represent
the upper and the down ports of the BS, respectively.

With our faithful single-photon transmission method (see Method),
Alice and Bob can share  photon pairs in a maximally entangled
state, shown in Fig. 5. In detail, after a photon pair from the
source passes through the two encoders, its state becomes
\begin{eqnarray}
|\varphi\rangle_{ab_1}\!\!&=&\!\!\frac{1}{2\sqrt{2}}|h\rangle_a|h\rangle_b\otimes
[(|u_l\rangle_a\!+\!|d_l\rangle_a)\otimes(|u_l\rangle_b\!+\!|d_l\rangle_b)\nonumber
\\&&
+(|u_s\rangle_a\!-\!|d_s\rangle_a)\otimes(|u_s\rangle_b\!-\!|d_s\rangle_b)].
\end{eqnarray}
As the two photons \emph{a} and \emph{b} suffer from independent
collective noises  from the two channels, the influence of the
channels on the two photons can be described with two unitary
rotations $U^a_{C}$ and $U^b_{C}$ as follows:
\begin{eqnarray}
U^a_{C}|h\rangle &\xrightarrow[]{noise}& \delta_a|h\rangle+\eta_a|v\rangle,\\
U^b_{C}|h\rangle &\xrightarrow[]{noise}&
\delta_b|h\rangle+\eta_b|v\rangle,
\end{eqnarray}
where  $|\delta_i|^2+|\eta_i|^2=1$ ($i=a,b$). The influence on the
polarization of the photons arising from the channel noises can be
totally converted into that on the spatial DOF. The state of the
photons \emph{a} and \emph{b} arriving at Alice and Bob becomes
\begin{eqnarray}
|\varphi\rangle_{ab_2}&=&\frac{1}{\sqrt{2}}(|h\rangle_a|v\rangle_b+|v\rangle_a|h\rangle_b) \nonumber\\
&&\otimes(\delta_a|a_1\rangle+\eta_a|a_2\rangle)\otimes(\delta_b|b_1\rangle+\eta_b|b_2\rangle) \nonumber\\
&= &|\varphi\rangle_{ab}^p\otimes|\varphi\rangle_{ab}^s.
\end{eqnarray}
This is a two-photon Bell state $|\varphi\rangle_{ab}^p =
\frac{1}{\sqrt{2}}(|h\rangle_a|v\rangle_b+|v\rangle_a|h\rangle_b)$
in the polarization DOF of the photon pair $ab$. Simultaneously, it
is  a separable  superposition state $|\varphi\rangle_{ab}^s =
(\delta_a|a_1\rangle+\eta_a|a_2\rangle)\otimes(\delta_b|b_1\rangle+\eta_b|b_2\rangle)$
in the spatial DOF.

To entangle the stationary atomic ensembles $E_A$ and $E_B$, which
are initialized to be
$|\phi_{_{A}}\rangle=\frac{1}{\sqrt{2}}(|G\rangle_{_{A}}+|S\rangle_{_{A}})$
and
$|\phi_{_{B}}\rangle=\frac{1}{\sqrt{2}}(|G\rangle_{_{B}}+|S\rangle_{_{B}})$,
only two CPF gates are required if  Alice and Bob have  shared some
photon pairs in the Bell state $|\varphi\rangle_{ab}^p$. Let us take
the case that the photons $a$ and $b$ come  from the spatial modes
$a_2$ and $b_2$ as an example to detail the entanglement creation
process. As for the other cases,  the same entanglement between
$E_A$ and $E_B$ can be obtained  by a similar procedure  with or
without some single-qubit operations.

First, the photon \emph{a} suffers a Hadamard operation by passing
through  a half-wave plate \emph{H}. Second,  it is reflected by the
cavity or the mirror $M$, which is used to complete  the  CPF  gate
on  the photon \emph{a} and the ensemble $E_A$.  Third,  Alice
performs another  Hadamard operation on the photon \emph{a}. Now,
the  state of the composite system composed of the photons  \emph{a}
and \emph{b} and the ensembles $E_A$ and $E_B$ evolves into
$|\Phi\rangle_{PE_1}$,
\begin{eqnarray}
|\Phi\rangle_{PE_1}&=&\frac{1}{2}
\big[|h\rangle_{a}(|v\rangle_{b}|S\rangle_{_{A}}-|h\rangle_{b}|G\rangle_{_{A}}) \nonumber\\
&&+|v\rangle_{a}(|h\rangle_{b}|S\rangle_{_{A}}-|v\rangle_{b}|G\rangle_{_{A}})\big]\otimes|\varphi\rangle_{_{B}}.\;\;\;\;
\end{eqnarray}
Fourth, Alice measures the polarization state of the photon $a$ with
a setup composed of PBS and single-photon detectors $D_h$ and $D_v$.
If an $|h\rangle$ polarized photon is detected, the hybrid system
composed of \emph{b}, $E_A$, and $E_B$ will be projected into
\begin{eqnarray}
|\Phi\rangle_{PE_2}=\frac{1}{\sqrt{2}}(|v\rangle_{b}|S\rangle_{_{A}}-|h\rangle_{b}|G\rangle_{_{A}})\otimes|\varphi\rangle_{_{B}}.
\end{eqnarray}
If a $|v\rangle$ polarized photon is detected, the remaining hybrid
system can also be  transformed into the state $|\Phi\rangle_{PE_2}$
by a bit-flip operation
$\hat{\sigma}^{A}_{x}=|S\rangle_{_{A}}\langle{}G|+|G\rangle_{_{A}}\langle{}S|$
on the ensemble $E_A$.

Up to now, the original entanglement of the photon pair \emph{ab} is
mapped to the hybrid entanglement between the photon \emph{b} and
the ensemble $E_A$. In order to create the entanglement between
$E_A$  and $E_B$, Bob just performs the same operations as Alice
does. In brief,  before and after the CPF operation on the photon
\emph{b} and the ensemble $E_B$,  Bob performs two local Hadamard
operations on the photon \emph{b} with \emph{H}.  These operations
result in the entanglement between the photon \emph{b} and the two
atomic ensembles. The state $|\Phi\rangle_{PE_2}$ is changed into
$|\Phi\rangle_{PE_3}$
\begin{eqnarray}
|\Phi\rangle_{PE_3}&=&\frac{1}{2}\big[|v\rangle_{b}\otimes(|S\rangle_{_{A}}|S\rangle_{_{B}}+|G\rangle_{_{A}}|G\rangle_{_{B}}) \nonumber\\
&&
-|h\rangle_{b}\otimes(|G\rangle_{_{A}}|S\rangle_{_{B}}+|S\rangle_{_{A}}|G\rangle_{_{B}})\big].\;\;
\end{eqnarray}
If the detector D$_{h}$ at Bob's node is clicked, the  state of the
system composed of  $E_A$ and $E_B$ will be collapsed into the
desired entangled state
\begin{eqnarray}
|\Psi\rangle_{_{AB}}=\frac{1}{\sqrt{2}}(|G\rangle_{_{A}}|S\rangle_{_{B}}+|S\rangle_{_{A}}|G\rangle_{_{B}}).
\end{eqnarray}
As for the case that the photon \emph{b} is in the state
$|v\rangle$, they can also obtain the desired entangled state
$|\Psi\rangle_{_{AB}}$ with an additional bit-flip operation
$\hat{\sigma}^B_{x}$ on $E_B$.

\subsection{Entanglement swapping on atomic ensembles with a PCG.}

After the parties produce successfully the entanglement between each
two atomic ensembles in the neighboring nodes, they can extend the
entanglement to a further distance by entanglement swapping. Let us
use the case with three nodes as an example to describe the
principle for connecting the two non-neighboring nodes.

\begin{figure}[h]
\begin{center}
\includegraphics[width=8 cm]{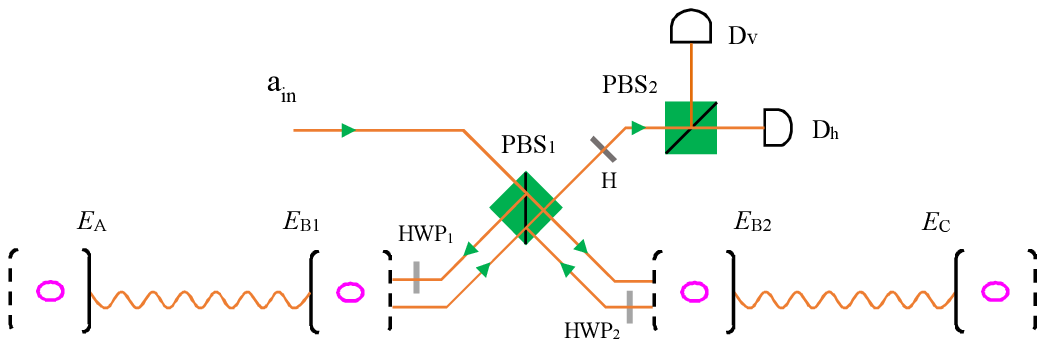}
\caption{(Color online) Schematic setup for entanglement swapping
with the simplified PCG. } \label{Fig6}
\end{center}
\end{figure}

Suppose the atomic ensembles $E_A$ and $E_C$ belong to  the two
non-neighboring nodes Alice and Charlie, respectively, and the two
ensembles $E_{B_1}$ and $E_{B_2}$ belong to the middle node Bob,
shown in Fig. 6.  The two ensembles $E_AE_{B_1}$ are in the state
$|\Psi\rangle_{_{AB_1}}=\frac{1}{\sqrt{2}}(|G\rangle_{_{A}}|S\rangle_{_{B_1}}-|S\rangle_{_{A}}|G\rangle_{_{B_1}})$
and the two ensembles $E_{B_2}E_C$ are in the state
$|\Psi\rangle_{B_2C}=\frac{1}{\sqrt{2}}(|G\rangle_{_{B_2}}|S\rangle_{_{C}}+|S\rangle_{_{B_2}}|G\rangle_{_{C}})$.
After a parity-check measurement  performed on the two local
ensembles $E_{B_1}$ and $E_{B_2}$  with a PCG shown in Fig. 3 (b),
the state of the system composed of the four ensembles $E_A$, $E_C$,
$E_{B_1}$,  and $E_{B_2}$  evolves into an entangled one.  If the
outcome of the parity-check measurement on the ensembles $B_1B_2$ is
odd,   the composite  system composed of  $E_{B_1}$, $E_{B_2}$,
$E_A$, and $E_{_C}$ will be projected into the state
\begin{eqnarray}
|\Psi\rangle_{E}&\!\!=\!\!&\frac{1}{\sqrt{2}}(|G\rangle_{_{B_1}}|S\rangle_{_{B_2}}|S\rangle_{_{A}}|G\rangle_{_{C}}
\!+\!|S\rangle_{_{B_1}}|G\rangle_{_{B_2}}|G\rangle_{_{A}}|S\rangle_{_{C}}),\nonumber\\
\end{eqnarray}
which is a four-qubit Greenberger-Horne-Zeilinger state. The
decoherence of both $E_{B_1}$ and $E_{B_2}$ has an awful influence
on the system composed of $E_{A}$ and $E_{C}$ as it decreases the
fidelity of the entanglement of the system. In order to disentangle
the two ensembles $E_{B_1}$ and $E_{B_2}$ from the system, the party
at  the middle node could first perform a Hadamard operation on the
two ensembles and then apply a parity-check measurement on them.  If
the outcome of the second parity-check measurement is even,  the
composite  system composed of the four ensembles  $E_{B_1}$,
$E_{B_2}$, $E_A$, and $E_C$  is projected into the state
\begin{eqnarray}
|\Psi\rangle_{E'}\!&=&\!\frac{1}{2}(|G\rangle_{_{B_1}}|G\rangle_{_{B_2}}+|S\rangle_{_{B_1}}|S\rangle_{_{B_2}})\nonumber\\
&&
\otimes(|S\rangle_{_{A}}|G\rangle_{_{C}}+|G\rangle_{_{A}}|S\rangle_{_{C}}),
\end{eqnarray}
where the ensembles $E_{B_1}$ and $ E_{B_2}$ are decoupled   from
the system composed of the two nonlocal ensembles $E_{A}$ and
$E_{C}$ which are in the maximally entangled state
$|\Psi\rangle_{AC}=\frac{1}{\sqrt{2}}(|G\rangle_{_{A}}|S\rangle_{_{C}}+|S\rangle_{_{A}}|G\rangle_{_{C}})$.

In the discussion above, we use the outcomes (odd, even) of the two
successive parity-check measurements as an example to describe the
principle of the entanglement swapping between the four atomic
ensembles. In fact, the other cases that the outcomes of each
parity-check measurement is either an odd one or an even one can
also be used for the entanglement swapping with only a single-qubit
operation on the ensemble  $E_A$, shown in Table. 1.

\begin{table}
\centering\caption{The relation between the single-qbuit operation
on the ensemble $E_{A}$ for entanglement swapping and the outcomes
of the parity-check measurements on the two atomic ensembles at the
middle node.  $P_1$ and $P_2$ denote the outcomes of the first and
the second parity-check measurements. Here $\hat{\sigma}_I=\vert
G\rangle_A\langle G\vert + \vert S\rangle_A\langle S\vert$,
$\hat{\sigma}_z=\vert G\rangle_A\langle G\vert - \vert
S\rangle_A\langle S\vert$, $\hat{\sigma}_y=\vert G\rangle_A\langle
S\vert - \vert S\rangle_A\langle G\vert$, and $\hat{\sigma}_x=\vert
G\rangle_A\langle S\vert +\vert S\rangle_A\langle G\vert$. }
\begin{tabular}{ccc}
\hline\hline
\quad{}\quad{}P$_1$ \quad\quad&   \quad{}\quad{}P$_2\quad\quad$ & \quad{}\quad{}E$_{A}\quad\quad$ \\
\hline
$v$  & $h$  & $\hat{\sigma}_I$ \\
$v$  & $v$  & $\hat{\sigma}_z$ \\
$h$  & $v$  & $\hat{\sigma}_y$ \\
$h$  & $h$  & $\hat{\sigma}_x$ \\
\hline\hline
\end{tabular}\label{tab1}
\end{table}

\section{Discussion}

We would like to briefly discuss the imperfections of our quantum
repeater protocol. The photon loss is the main imperfection, which
is also of crucial importance for the previous quantum repeaters
with photon interference \cite{high-frep1,qdrepeater1,qdrepeater12,
qdrepeater2,qdrepeater3,qdions,atomrepeater2,reviewqr,DLCZ,QRsimontimemul,
QRsingleSangouard,QRsimonspatialmul,QRzhao,QRchen,QRdoubleSangouard,QRlingw,QRxiongsj}.
The photon loss happens, due to the fiber absorbtion, diffraction,
the cavity imperfection, and the inefficiency of the single-photon
detectors. It will decrease the success probability and prolong the
time needed for establishing the quantum repeater. Since the memory
node in this protocol is implemented with the atomic ensemble, the
local operation between two collective quantum states $|G\rangle$
and $|S\rangle$ of the memory node, can be performed with collective
laser manipulations \cite{QRBrion}, while excitations of
higher-order collective states can be suppressed efficiently with
the  Rydberg blockade \cite{Rydbergcollectiveencodeing}. During the
entanglement swapping process, to detect the collective state of two
ensembles in the centering   nodes, fluorescent detection
\cite{floures} can be used, since the detection efficiencies of
$99.99\%$ for trapped ions have been experimentally demonstrated
\cite{floures_ion}. Moreover, with the current significant progress
on the source of entangled photon pairs, the repetition rate as high
as $10^6/10^7S^{-1}$ has been achieved \cite{fwm}, so our
entanglement distribution process can be performed with a high
efficiency.

In summary, we have proposed a high-efficiency quantum repeater with
atomic ensembles embedded in optical cavities as the memory nodes,
assisted by single-photon faithful transmission. By encoding the
polarization qubit into the time-bin qubit,  our faithful
single-photon transmission can be completed with only linear-optical
elements, and neither time-slot discriminator nor  fast PCs is
required \cite{Faithful,QRgaom,QRdfs,QRzhangbb,Faithfulerr}. The
heralded  entanglement creation between the neighboring nodes is
achieved with a CPF gate between the atomic ensemble and the photon
input in each node, which makes our scheme more convenient than the
one with post selection \cite{QRBrion}, although both efficiencies
of our quantum repeaters are identical and maximal among all the
exciting quantum repeater schemes when multi-mode speed up is not
considered \cite{QRsimontimemul,QRsimonspatialmul}. Besides, no
additional classical information is involved to determinate the
state of the entangled atomic ensembles, since the parties can
create a deterministic entanglement up to a feedback upon the
results of photon detection. The quantum swapping process is
deterministically completed with a simplified PCG involving only one
input-output process,  which makes our scheme far more efficient
than the ones based on linear optical elements \cite{reviewqr}.

\section{Methods}

\subsection{Faithful single-photon transmission}

Our  protocol for  deterministic polarization-error-free
single-photon transmission can be details as follows.  Assuming the
initial state of the single photon to be transmitted  is
$|\varphi\rangle=\mu|h\rangle+\nu|v\rangle$ ($|\mu|^2+|\nu|^2=1$).
After passing through the encoder, the photon launched into the
noisy channel evolves into
\begin{eqnarray}
|\varphi'\rangle=\frac{1}{\sqrt{2}}|h\rangle\otimes(\nu|u_l\rangle+\nu|d_l\rangle+\mu|u_s\rangle-\mu|d_s\rangle),
\end{eqnarray}
where the subscripts \emph{l} and \emph{s} represent the photons
passing through the long path and short path of the encoder,
respectively. When the optical path difference between \emph{l} and
\emph{s} is small, the two time bins are so close that they suffer
from the same fluctuation from the optical fiber channels
\cite{QKD2,Qcryptography,Faithful,QRgaom,QRdfs,QRzhangbb,Faithfulerr,Qpsimon,Qdeterministicpurification,DEPP2,DEPP3,QpurificationLi,QpurificationShengSR}.
The noise of the channel can be expressed with a unitary
transformation U$_C$ as follows:
\begin{eqnarray}
U_C|h\rangle \xrightarrow[]{\text{  noise }}
\;\;\delta|h\rangle+\eta|v\rangle,
\end{eqnarray}
where $ |\delta|^2+|\eta|^2=1$. After the photon passes through the
channels, a $\pi$ phase shifter $P_{\pi}$ on the \emph{d} channel
is applied, and the state of the photon becomes
\begin{eqnarray}
|\varphi''\rangle= \frac{1}{\sqrt{2}}(\delta|h\rangle\!+\eta|v\rangle)\otimes(\nu|u_l\rangle\!-\!\nu|d_l\rangle+\mu|u_s\rangle+\!\mu|d_s\rangle).\nonumber\\
\end{eqnarray}
With a decoder  composed of a BS, an HWP, and a PBS, shown in Fig. 5
(b), the evolution  of the photon can be described as follows:
\begin{eqnarray}
|\varphi''\rangle \!\!&\xrightarrow[]{\text{ BS }}&\!\! (\delta|h\rangle\!+\!\eta|v\rangle)\otimes(\nu|d_{ls}\rangle\!+\!\mu|u_{sl}\rangle) \nonumber\\
 \!\! &\xrightarrow[]{\text{ HWP }}&\!\!
\nu|d_{ls}\rangle\otimes(\delta|h\rangle\!+\!\eta|v\rangle)
\!+\!\mu|u_{sl}\rangle\otimes(\delta|v\rangle\!+\!\eta|h\rangle) \nonumber\\
 \!\!&\xrightarrow[]{\text{  PBS }}&\!\!
\delta|a_1\rangle(\nu|h\rangle\!+\!\mu|v\rangle)\!+\!\eta|a_2\rangle(\mu|h\rangle\!+\!\nu|v\rangle) \nonumber\\
\!\!& \xrightarrow[]{\text{  HWP }}&\!\!
(\mu|h\rangle\!+\!\nu|v\rangle)\otimes(\delta|a_1\rangle\!+\!\eta|a_2\rangle).
\end{eqnarray}
Here the subscripts $ls$ ($sl$) represent the photon that passes
through the long (short) path of the encoder and the short (long)
path of the decoder, respectively. The difference between the long
path and the short one for the encoder is designed to be the same as
that for the decoder.  Without any time-slot discriminator,  one can
get the error-free photon in either  the   output  $a_1$ or $a_2$ at
a deterministic time slot.

\subsection{Performance of  CPF  and  PCG with current experimental parameters.}

Before we analyze the fidelity of the quantum entanglement
distribution and entanglement swapping in our quantum repeater
scheme, we first discuss the practical performance of  the CPF gate
and the PCG based on the recent experiment advances
\cite{gvarious,fibercavity1,fibercavity2}. We define the fidelity of
a quantum process (or a quantum gate)  as $F = |\langle \Psi_i
|\Psi_r \rangle|^2$, where $\vert \Psi_i\rangle$ and $\vert
\Psi_r\rangle$ are the output states of the quantum system in the
quantum process  (or the quantum gate) in the ideal condition and
the realistic condition, respectively \cite{reviewqr}.

By combining a fibre-based cavity with the atom-chip technology,
Colombe \emph{et al}. \cite{gvarious} demonstrated the strong
atom-field coupling in a recent experiment in which each $^{87}$Rb
atom in Bose-Einstein condensates is identically and strongly
coupled to the cavity mode. In this experiment, all the atoms are
initialized to be the hyperfine zeeman state $|5S_{1/2}, F=2,
m_f=2\rangle$. The dipole transition of $^{87}$Rb $|5S_{1/2},
F=2\rangle$ $\mapsto$ $|5P_{3/2}, F'=3\rangle$ is resonantly coupled
to the cavity mode with the maximal single-atom coupling strength
$g_0=2\pi\times 215$ MHz. Meanwhile, the cavity photon decay rate is
$\kappa=2\pi\times53$ MHz and the atomic spontaneous emission rate
of $|5P_{3/2}, F'=3\rangle$ is $\gamma=2\pi\times3$ MHz. The
whispering-gallery microcavities (WGMC)  \cite{WGMC1} might be
another potential experimental realization of our scheme. The
parity-time-symmetry breaking is realized in a system of two
directly coupled WGMC \cite{WGMC2} and the controlled loss is also
achieved with WGMC \cite{WGMC3}, which enables the on-chip
manipulation and control of light propagation. In addition, the
routing of single photons has been demonstrated by the atom-WGMC
coupled unit controlled by a single photon \cite{WGMC4}.

Under an ideal condition,  the reflection coefficients  of the
input-output processes are  $r_{_0}{(\delta')}\simeq-1$ and
$r{(\delta')}\simeq1$.  In this time,  the input $|h\rangle$
polarized photon \emph{a} will get a $\pi$ phase shift when the
embedded atomic ensemble $E_{_A}$ is in the state $|G\rangle$;
otherwise, there is  no phase shift on the photon $a$.  The fidelity
of both the CPF gate (shown in Fig.4 (a)) and the PCG (shown in Fig.
4 (b)) can reach unity. In a realistic atom-cavity system, the
relationship between the input and output field is outlined in Eqs.
(\ref{r}) and (\ref{r0}). In this time, after the party operates the
photon \emph{a} and the ensemble $E_{_A}$ with the CPF gate, the
output state of the composite system becomes
\begin{eqnarray}
\begin{split}
|\Phi'\rangle_{_Ep}\;=\;&\frac{1}{\sqrt{C}}[\mu'|G\rangle\otimes(r_0\mu|h\rangle+\nu|v\rangle)\\
&+\nu'|S\rangle\otimes(r \mu|h\rangle+\nu|v\rangle)].
\end{split}
\end{eqnarray}
Here the normalized coefficient
$C=|r_0\cdot\mu'\cdot\mu|^2+|r\cdot\nu'\cdot\mu|^2+|\mu'\cdot\nu|^2+|\nu'\cdot\nu|^2.$
The fidelity of the CPF gate
$F_{cpf}=|_{_Ep}\langle\Phi|\Phi'\rangle{_{_Ep}}|^2$ depends on the
input state of the system composed of the photon and the atomic
ensemble. In the symmetric case with $\mu=\mu=\mu'=\nu'=1/\sqrt{2}$,
the fidelity $F_{cpf}$ can be simplified to be
\begin{eqnarray}
F_{cpf}=\!\frac{1}{4}+\frac{1-\!Re(r\cdot{}r_0^{\ast})-\!2Re(r_0)+\!2Re(r)}{2(2+|r|^2+|r_0|^2)}.
\end{eqnarray}
Meanwhile, the efficiency $\eta_{cpf}$ of the CPF gate, which is
defined as the probability that the photon  clicks either detectors
after being reflected by the CPF gate, can be detailed as
\begin{eqnarray}
\eta_{cpf}= \frac{1}{2} + \frac{|r|^2 + |r_0|^2}{4}.
\end{eqnarray}

In a realistic condition,  the output state of the composite system
composed of \emph{a}, $E_A$, and $E_B$  in the PCG  process before
the single photon is detected becomes
\begin{eqnarray} 
|\Phi''\rangle_{p_{AB}}&=&\frac{1}{\sqrt{C}}\{|v\rangle(r_0-r)(|G,S\rangle-|S,G\rangle) \nonumber\\
&&+|h\rangle[\sqrt{2}r_0|G,G\rangle+\sqrt{2}r_0|S,S\rangle \nonumber\\
&& +(r_0+r)(|G,S\rangle+|S,G\rangle)]\}.
\end{eqnarray}
Compared with the ideal output state described in Eq.(\ref{phipab}),
if an $|h\rangle$ polarized photon is detected, the fidelity of the
PCG gate $F_{pcg}$ can be expressed as
\begin{eqnarray}
F_{pcg}=\!\frac{|r|^2+\!|r_0|^2\!-\!2Re(r\cdot{}r_0^{\ast})}{3(|r|^2+|r_0|^2)+2Re(r\cdot{}r_0^{\ast})}.
\end{eqnarray}
When the photon in the state $|v\rangle$ is detected, the fidelity
of the PCG is $F'_{pcg}=1$.  The success of the PCG is heralded when
a single photon is detected after the parity-check process, no
matter what the state the photon evolves to be. The efficiency
$\eta_{pcg}$ of the PCG process can be defined as the probability
that the probe photon is detected after it is reflected by the two
cavities,  that is,
\begin{eqnarray}
\eta_{cpf}= \frac{|r|^2+|r_0|^2}{2}.
\end{eqnarray}

Since the absolute value of the relative phase shift during the
input-output process depends on the frequency of the input photon,
it  decreases  smoothly  with  the detuning  $\delta'$  between  the
input photon and the cavity mode, shown in Fig. 3.

\begin{figure}[!h]
\begin{center}
\includegraphics[width=8.0cm]{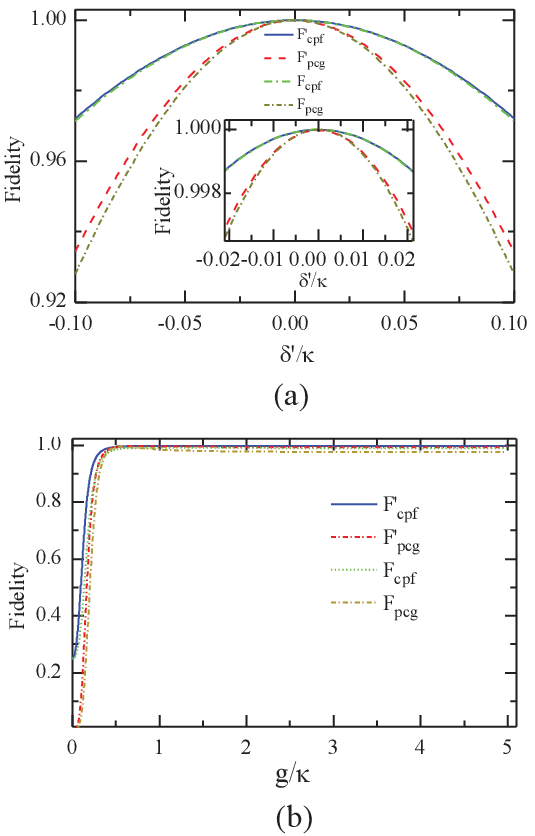}%
\caption{(Color online) (a) Fidelities of CPF gate and PCG vs the
scaled detuning. $F'_{cpf}$ and $F'_{pcg}$ is performed with the
scaled coupling rate $g/\kappa=2.0283$ and $\gamma/\kappa=0.0566$,
$F_{cpf}$ and $F_{pcg}$ are performed with the scaled coupling rate
$g/\kappa=4.0566$ and $\gamma/\kappa=0.0566$ \cite{gvarious}. (b)
Fidelities of CPF gate and PCG gate VS the scaled coupling rate.
$F'_{cpf}$ and $F'_{pcg}$ are performed with the scaled detuning
$\delta'/\kappa=0.0283$ and $\gamma/\kappa=0.0566$, $F_{cpf}$ and
$F_{pcg}$ is performed with   $\delta'/\kappa =
\gamma/\kappa=0.0566$.}\label{Fig7}
\end{center}
\end{figure}

\begin{figure}[!h]
\begin{center}
\includegraphics[width=8 cm]{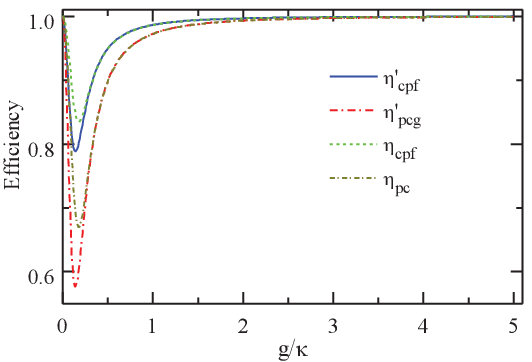}
\caption{(Color online)  Efficiencies of CPF gate and PCG gate vs
the scaled coupling rate. $\eta'_{cpf}$ and $\eta'_{pcg}$ is
performed with the scaled detuning $\delta'/\kappa=0.0283$ and
$\gamma/\kappa=0.0566$, $\eta_{cpf}$ and $\eta_{pcg}$ are performed
with  $\delta'/\kappa = \gamma/\kappa=0.0566$. } \label{Fig8}
\end{center}
\end{figure}

The fidelity of the CPF gate $F_{cpf}$ changes with the detuning
$\delta'$, shown in Fig. 7(a). Here the parameters are chosen as
$g/\kappa=2.0283$ or $4.0566$ and $\gamma/\kappa=0.0566$
 \cite{gvarious}. When the linewidth of the input photon is
$\delta=2|\delta'|_{max}$ with the maximal detuning
$|\delta'|_{max}=0.5\gamma$ ($\gamma$),   $F_{cpf}$ is larger than
$F_{cpf}(|\delta'|_{max})=0.9974$ ($0.9906$) for $g/\kappa=4.0566$.

The fidelity of the PCG depends on the coupling rate $g/\kappa$, as
shown in Fig. 7(b)  with the detuning $|\delta'|_{max}=0.5\gamma$ or
$\gamma$. When the maximal detuning
 of the input photon is  $|\delta'|_{max}=0.5\gamma$,
the high-performance parity-check gate can be achieved with the
fidelity $F_{pcg}$ higher than $F_{pcg}(|\delta'|_{max})=0.9944$ and
$0.9938$ for $g/\kappa=2.0283$ and $g/\kappa=4.0566$, respectively.

The efficiencies of the CPF gate and the PCG process versus the
coupling rate $g/\kappa$ are shown in Fig. 8. When the bandwidth of
the probe photon is on the scale of  $\gamma$, both efficiencies
$\eta_{cpf}$ and $\eta_{pcg}$ are robust to the variation of
$g/\kappa$ with the  parameters above \cite{gvarious}. 
In detail, when the maximal detuning $|\delta'|_{max}$ of the input
photon is less than $0.5\gamma$, $\eta_{cpf}$ and $\eta_{pcg}$ are
higher than  $0.9966$ and $0.9932$, respectively. When
 $|\delta'|_{max}=\gamma$, $\eta_{cpf}=0.9991$
and $\eta_{pcg}=0.9983$ are achievable.

\begin{figure}[!h]
\begin{center}
\includegraphics[width=8 cm]{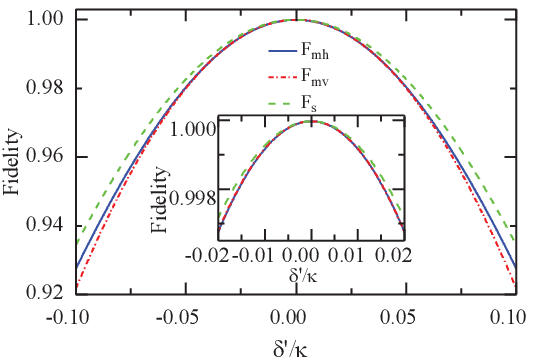}
\caption{(Color online)\label{Fig9}  Fidelities of F$_{mh}$,
F$_{mv}$ and F$_s$ \emph{vs} the detuing, with the scaled coupling
rate $g/\kappa=2.0283$ and $\gamma/\kappa=0.0566$.} \label{Fig8}
\end{center}
\end{figure}

\begin{figure}[!h]
\begin{center}
\includegraphics[width=8cm]{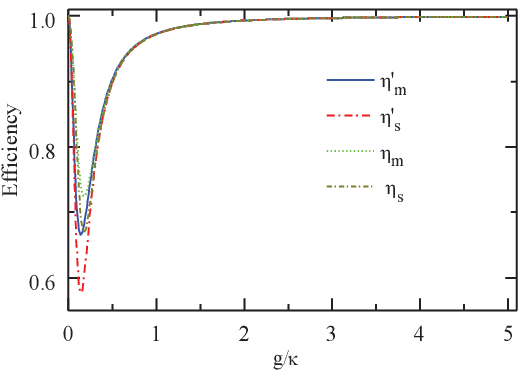}
\caption{(Color online)  Efficiencies of entanglement distribution
and entanglement swapping processes vs the scaled coupling rate.
$\eta'_{m}$ and $\eta'_{s}$ is performed with the scaled detuning
$\delta'/\kappa=0.0283$ and $\gamma/\kappa=0.0566$, $\eta_{m}$ and
$\eta_{s}$ are performed with  $\delta'/\kappa =
\gamma/\kappa=0.0566$. } \label{Fig10}
\end{center}
\end{figure}

\subsection{Performance of entanglement distribution and entanglement swapping.}

Now, let us discuss the fidelities and the efficiencies of the
entanglement distribution and entanglement swapping in our quantum
repeater scheme.  After Alice performs  the local operations on the
photon \emph{a} and detects an $|h\rangle$ polarized photon, the
composite  system composed  of the photon \emph{b} and the ensembles
$E_A$ and $E_B$  will be projected into the state
$|\Phi'_{PE_2}\rangle$, instead of $|\Phi\rangle_{PE_2}$. Here
\begin{eqnarray}
|\Phi'\rangle_{PE_2}&=&\frac{1}{\sqrt{C}}
[(r_0-1)|h\rangle\otimes|G\rangle_{_{A}} \nonumber\\
&&+(r_0+1)|v\rangle\otimes|G\rangle_{_{A}}+(r-1)|h\rangle\otimes|S\rangle_{_{A}} \nonumber\\
&&+(r+1)|v\rangle\otimes|S\rangle_{_{A}}]\otimes|\varphi\rangle_{_{B}},
\end{eqnarray}
where the normalized coefficient
$C=2[|r_0-1|^2+|r_0+1|^2+|r-1|^2+|r+1|^2]$. And then, the same
operations, i.e.,  a CPF gate sandwiched by two Hadamard operations,
are performed by Bob on the photon \emph{b}.  After these
operations, the state of the composite  system composed  of  the
photon \emph{b} and the two ensembles $E_A$ and $E_B$ evolves into
\begin{eqnarray}
|\varphi\rangle_{pE_2}&=&\frac{1}{\sqrt{C'}} \{|h\rangle\!\otimes\![(r_0^2-1)|G\rangle_{_{A}}\!\otimes\!|G\rangle_{_{B}} \nonumber\\
&& +(r_0\cdot{}r-1)(|G\rangle_{_{A}}\!\otimes\!|S\rangle_{_{B}}\!+\!|S\rangle_{_{A}}\!\otimes\!|G\rangle_{_{B}}) \nonumber\\
&& +(r^2-1)|S\rangle_{_{A}}\otimes|S\rangle_{_{B}}] \nonumber\\
&& +|v\rangle\!\otimes\![(r_0^2+1)|G\rangle_{_{A}}\!\otimes\!|G\rangle_{_{B}} \nonumber\\
&& +(r_0\cdot{}r+1)(|G\rangle_{_{A}}\!\otimes\!|S\rangle_{_{B}}\!+\!|S\rangle_{_{A}}\!\otimes\!|G\rangle_{_{B}}) \nonumber\\
&& +(r^2+1)|S\rangle_{_{A}}\otimes|S\rangle_{_{B}}]\},
\end{eqnarray}
where the normalized coefficient
$C'=2[|r_0^2|^2+|r^2|^2+2|r\cdot{}r_0|^2+4]$. One can obtain the
fidelity of the entanglement distribution process $F_{mh}$ and
$F_{mv}$ for the cases that $D_h'$ and $D_v'$ at the Bob's node are
clicked, respectively.
\begin{eqnarray}
F_{mh}&=&\frac{2|r\cdot{}r_0-1|^2}{|r_0^2-1|^2+|r^2-1|^2+2|r\cdot{}r_0-1|^2}, \nonumber\\
F_{mv}&=&\frac{\frac{1}{2}|r^2+r_0^2+2|^2}{|r_0^2+1|^2+|r^2+1|^2+2|r\cdot{}r_0+1|^2}.\;\;\;\;\;
\label{fmhfmv}
\end{eqnarray}
If one defines the efficiency $\eta^h_m$  as the probability  that
Alice detects an $|h\rangle$ polarized photon while Bob detects a
photon in either $|h\rangle$ or $|v\rangle$ polarization, one has
\begin{eqnarray}
\eta^h_{m}=\frac{C}{32}\cdot\frac{C'}{C}=\frac{|r_0^2|^2+|r^2|^2+2|r\cdot{}r_0|^2+4}{16}.
\end{eqnarray}
In the above discussion, we detail the performance of our
entanglement distribution conditioned on the detection of an
$|h\rangle$ polarization photon at Alice's node. Considering the
symmetric property of the system, one can easily obtain the
performance of the entanglement distribution upon the detection of a
$|v\rangle$ polarization photon at Alice's node.  Now, the
fidelities $F'_{mh}$ and $F'_{mv}$ for the cases that $D_h'$ and
$D_v'$ are clicked at Bob's node, have the following relations to
that for the cases that an $|h\rangle$ polarized photon is detected
by Alice, $F'_{mh}=F_{mv}$ and $F'_{mv}=F_{mh} $, see Eq.
(\ref{fmhfmv}) for detail. Meanwhile, the efficiency $\eta^v_m$ of
the entanglement distribution process when Alice detects a photon in
$|v\rangle$ polarization is identical to  $\eta^h_m$.  The total
efficiency $\eta_m$ of the  entanglement distribution can be written
as
\begin{eqnarray}
\eta_m=\eta^h_{m}+\eta^v_{m}=\frac{|r_0^2|^2+|r^2|^2+2|r\cdot{}r_0|^2+4}{8}.
\end{eqnarray}

In our entanglement swapping process, two PCGs are applied on the
two ensembles $E_{B_1}$ and $E_{B_2}$ at the middle node. In fact,
only one PCG is enough if a  single-atomic-ensemble measurement  on
each of the two ensembles $E_{B1}$ and $E_{B2}$ is utilized after
the local Hadamard operations. After these measurements, the system
composed of the two remote ensembles $E_A$ and $E_C$ is in the state
$|\Psi\rangle_{_{AC}}$ with or without a local unitary operation.
When the fluorescent measurement \cite{floures} or field-ionizing
the atoms \cite{QRzhaohan} with the help of Rydberg excitation are
used, the state detection on atomic ensembles could be performed
with a near-unity efficiency.  In other words,  the fidelity of the
quantum entanglement swapping process can equal to that of the PCG
operation.

The fidelities of both the entanglement distribution and the
entanglement swapping in our repeater scheme are shown in  Fig. 9.
One can see that all $F_{mh}$, $F_{mv}$, and $F_{s}=F_{pcg}$ are
larger than $0.9936$  with  the  parameters ($g$, $\kappa$,
$\gamma$)$\;=2\pi\times$($215$, $53$, $3$) MHz achieved  in
experiment \cite{gvarious}. Meanwhile, all efficiencies involved in
our quantum repeater protocol, shown in Fig. 10, can be larger than
$0.9931$ when the effective coupling $g/\kappa>2.0283$ with
$\delta'/\kappa = \gamma/\kappa=0.0566$.  In a recent experiment
with a fiber-based Fabry\emph{-}Perot cavity constituted by CO$_2$
laser-machined mirrors \cite{fibercavity0}, the maximal coupling
strength as high as $g=2\pi\times2.8$ GHz is achieved for single Rb
atoms and the cavity decay rate is $\kappa=2\pi\times0.286$ GHz
$\simeq95\gamma$. In this time, $g/\kappa = 9.79$ is achieved, and a
better performance of our scheme is attainable.


\section*{ACKNOWLEDGEMENTS}

TL was supported by the China Postdoctoral Science Foundation under
Grant No.  2015M571011.
 FG was supported by the National Natural Science Foundation of
China under Grant Nos. 11174039 and 11474026, and the Open
Foundation of State key Laboratory of Networking and Switching
Technology (Beijing University of Posts and Telecommunications)
under Grant No. SKLNST-2013-1-13.

\end{document}